# Pre-conceptual Design Assessment of DEMO Remote Maintenance

A. Loving[a], O. Crofts[a], N. Sykes[a], D. Iglesias[a], M. Coleman[a], J. Thomas[a], J. Harman[b], U. Fischer[c], J. Sanz[d], M. Siuko[e], M. Mittwollen[f], et.al.

[a]EURATOM/Culham Center Fusion Energy, CulhamScience Centre OX14 3DB Abingdon, UK
[b]EFDA Close Support Unit Garching, Boltzmannstaße 2, D-85748, Garching Bei München, Germany
[c]Karlsruhe Institute of Technology, Institute for Neutron Physics and Reactor Technology, Hermann-von-Helmholtz-Platz 1, D-76344 Eggenstein-Leopoldshafen, Germany
[d]Instituto de Fusión Nuclear/UPM, Madrid, Spain
[e]VTT Technical Research Centre of Finland, P.O. Box 1000, FI-02044 VTT, Finland
[f]Karlsruhe Institute of Technology, Institut für Fördertechnik und Logistiksysteme, Gotthard-Franz-Straße 8, Geb.50.38, 76131 Karlsruhe, Germany

EDFA, as part of the Power Plant Physics and Technology programme, has been working on the pre-conceptual design of a Demonstration Power Plant (DEMO). As part of this programme, a review of the remote maintenance strategy considered maintenance solutions compatible with expected environmental conditions, whilst showing potential for meeting the plant availability targets. A key finding was that, for practical purposes, the expected radiation levels prohibit the use of complex remote handling operations to replace the first wall. In 2012/13, these remote maintenance activities were further extended, providing an insight into the requirements, constraints and challenges. In particular, the assessment of blanket and divertor maintenance, in light of the expected radiation conditions and availability, has elaborated the need for a very different approach from that of ITER. This activity has produced some very informative virtual reality simulations of the blanket segments and pipe removal that are exceptionally valuable in communicating the complexity and scale of the required operations. Through these simulations, estimates of the maintenance task durations have been possible demonstrating that a full replacement of the blankets within 6 months could be achieved. The design of the first wall, including the need to use sacrificial limiters must still be investigated. In support of the maintenance operations, a first indication of the requirements of an Active Maintenance Facility (AMF) has been elaborated.

## 1. Introduction

A demonstration fusion power station presents a number of challenges for maintenance. The high neutron flux anticipated in these power stations will rapidly degrade the plasma facing components, particularly the divertor and blanket modules so that they will require frequent replacement [1]. Furthermore, this neutron flux activates isotopes within these components and the surrounding structure, limiting the materials and equipment that will operate in the environment.

Further demand on the maintenance system is created by the fact that the cost of electricity is assessed to be inversely related to the availability of the machine [2], therefore a key driver for the design of maintenance systems is to make them as rapid as possible. It is also recognised that the maintenance strategy that is capable of meeting the stringent environmental conditions whilst striving to meet a high availability target for a power plant, will influence many aspects of the power plant design. It is therefore important for the maintenance strategy to be established early so as to have an equal input in the conceptual design process.

These requirements are balanced by the intrinsic machine design particularly [3]: minimisation of toroidal field ripple; structural support of the toroidal field coils; minimisation of the size and plasma proximity to the poloidal field coils; provision of tritium breeding; cooling; diagnostic service connections to the blanket modules and divertors; shielding of the superconducting coils and other systems from the neutron flux; and providing access for current drive and heating systems to the plasma.

## 2. Architectural evaluation of the DEMO Tokamak for maintainability

To resolve this, a broad review of architectures was conducted in 2011 to identify optimum architecture for maintenance. This examined the kinematics that would be most favourable for maintenance, the components that could be articulated to increase access potential and the optimum grouping of blanket module segments. The architectures considered were: the vertical maintenance system (VMS) proposed by Boccaccini and Nagy [4], with the segregation of the inboard and outboard segments on the mid-plane; the "NET" concepts proposed by Chazalon [5], where the access port takes advantage of maximum separation of the TF coils at 12° from the vertical; and two innovative concepts where the increased access space created by the removal of the divertor is exploited, one with the divertor located on the Tokamak floor with segments lowered through a floor mounted port and another with the divertor on the roof.

_author's email: antony.loving@ccfe.ac.uk_

The study concluded that the optimum architecture for maintenance was based around a vertical access extraction path [6-8], with sixteen upper vertical ports to give line of sight access to two inboard (IBS) and three outboard vertical module segments (OBS) [9], and sixteen equi-spaced lower divertor ports each to access three divertor cassettes [10].

## 3. Estimated radiation and decay heating effects

For a provisional DEMO model with HCLL-type blankets, global and local radiation fields have been calculated using both R2Smesh and R2S-UNED codes to check agreement between the codes and thereby validate the results [11,12]. Figure 1 below shows maps of the photon flux, where the divertor and blankets, and Tokamak components having been exposed to 1.57 and 6 full power years respectively. The figure shows the photon flux at one week and one year after the last plasma.

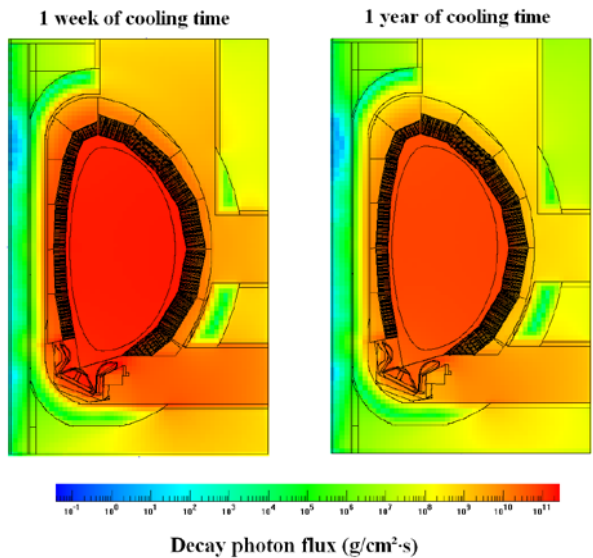

Fig. 1. Photon flux after 1 week and 1 year cooling time.

The high flux in the divertor port compared to the upper and equatorial ports is mostly due to neutron streaming between the outer blanket and the divertor. The geometry of this region will be changed to increase self-shielding and thereby minimise this effect.

The resultant maximum photon absorbed dose rates (in Gy/hr) of typical materials used by RHE is calculated as shown in table 1 for two locations and three durations after the last plasma.

Table 1. Calculated dose rates in Gy/hr

|  | 1 week | 1 month | 1 year |
|---|---|---|---|
| In-vessel | 2,300 | 1,500 | 800 |
| In a port | 95 | 80 | 15 |

### 3.1. Radiation tolerance studies

An assessment was made of the radiation tolerance of the components found in different types of remote handling equipment [13].

Dextrous manipulators were found to be the most sensitive. Using existing components they had a tolerance of less than 2MGy. This allows them to be used continuously in the port areas for the maintenance duration but only to be used in-vessel for short duration operations such as recovery, deployed on the multi-purpose deployer, see § 4.3.

## 4. Vertical maintenance system

The proposed vertical maintenance scheme shown in figure 2, requires the utilisation of all three types of port for remote operations:

### 4.1 Upper vertical port

The upper port is sized so as to allow the removal and installation of the multi-module segments (MMS) and to allow access to all the service connections. To enable multiple ports to be worked on at any one time remote handling equipment (RHE) casks have been designed to stay within the $1/16^{th}$ segment of the torus where they are operating. Before removing the MMS the divertor cassettes will need to be removed. This is compatible with the schedule maintenance strategy [14].

The RHE is deployed after removal of the bio-shield plug by the crane. The maintenance of the blanket modules involves several operations carried out by equipment deployed in four different types of cask [15]:

- An in-vessel mover (IVM) is installed through the divertor port for the disconnection of the lower supports of each MMS to the vacuum vessel and provides support during the initial MMS translations.

- A pipe joint handling cask (PJC) holds the equipment to cut, remove and store the pipe joints connecting the cooling and tritium breeding circuits to the MMS. The same cask is used for installation operations which include positioning, welding and NDT inspection.

- The blanket pipe sections between the port closure plate and the MMS will be maintained by the Port Closure Cask (PCC) equipped in a similar way to the PJC. The top of these pipes is clamped in groups of three and joined to the port flange by a set of bellows.

- The vertical maintenance crane (VMC) manipulates and lifts each blanket segment using a vertical maintenance transporter. It consists of three major parts: a static frame which is rigidly fixed to the cask, a telescopic frame which slides down within the static frame, and a crane mounted transporter which slides down within the telescopic frame.

The MMS and other in-vessel components require inspection, cleaning and minor unscheduled maintenance. Any failure or malfunction of the blankets would result in a complete substitution of up to five of the modules depending on which one is affected assuming there is no system capable of in-vessel repair.

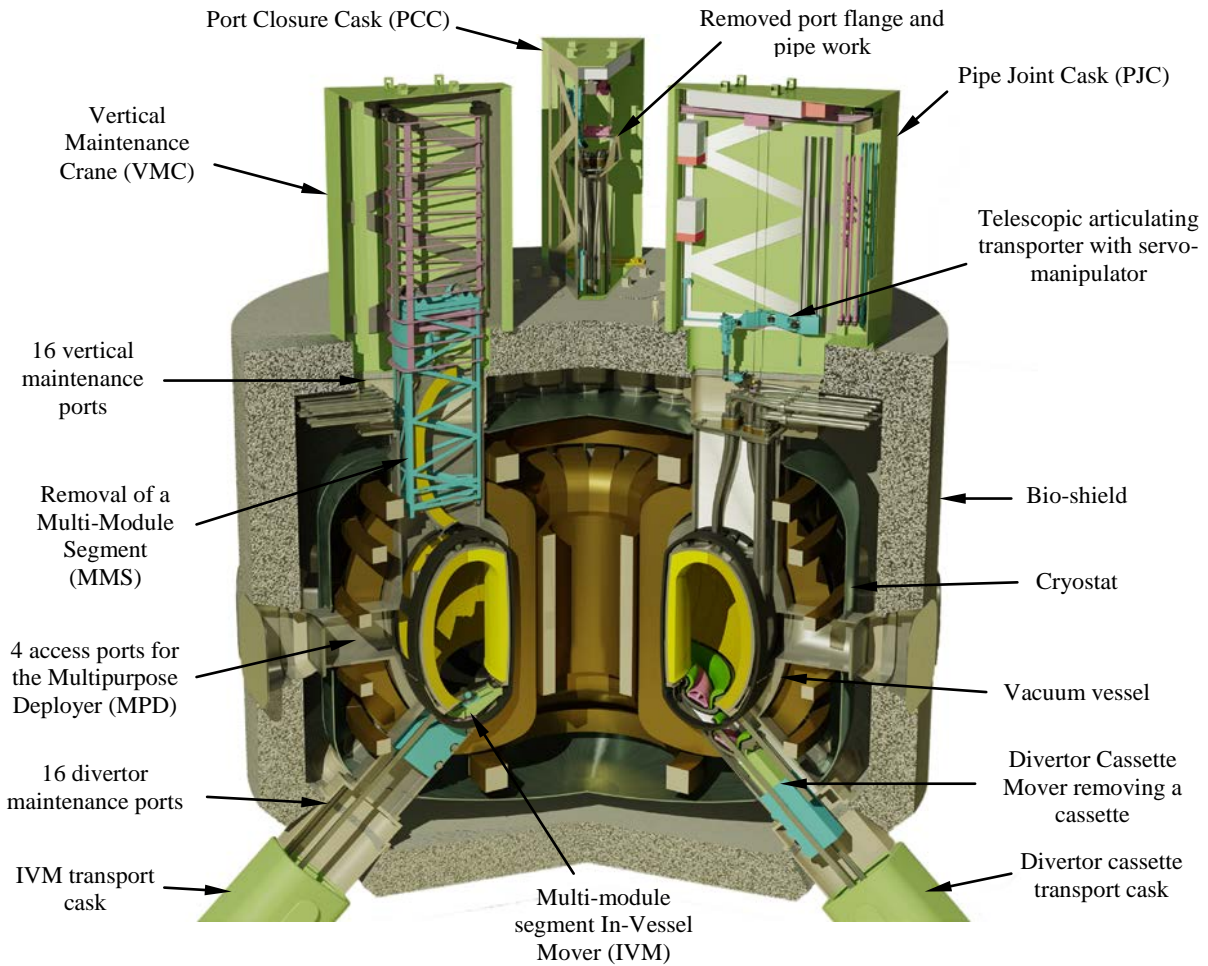

Fig. 2. Section of proposed DEMO Vertical Maintenance System architecture

### 4.2 Lower port

Lower ports are to be used for replacing the divertor cassettes and their services connections as well as the outlet pipe of the LiPb circuit in the case of the helium cooled lithium lead (HCLL) blanket option [10]. The port is orientated at 45° with the cask docking in such a way to allow a guided telescopic arm to reach the divertor cassette. This concept allows the majority of the RHE to remain outside of the high radiation areas.

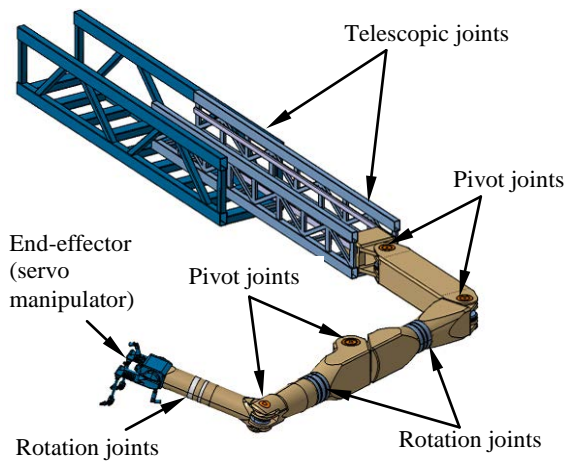

Fig 3. Multipurpose deployer

### 4.3 Equatorial ports

The equatorial port and blanket openings allow access for a multipurpose deployer (MPD) similar to that proposed for ITER, figure 3. The MPD has been dimensioned to carry out the tasks required to avoid the risk of lengthy plant downtimes due to minor issues. It would also be able to inspect and identify the origin of any problem inside the vacuum vessel, apply corrective actions, and also be available for recovery of the primary RHE used during first wall exchanges. The current proposal fits through the DEMO equatorial 1.45m square port [16] and has a reach of 90° in a toroidal direction. Substantiation analyses have been carried out considering a 0.6 ton payload.

## 5. Active Maintenance Facility

The Active Maintenance Facility (AMF), see figure 4, has been developed to support the VMS as described in § 4. The concept is for an extensive facility at 737,000m$^3$ that is able to meet the operational demands of the maintenance program [17]. It is necessary for the AMF to provide adequate component storage areas to cope with one full set of activated components and one replacement set. The full set of in-vessel components consists of 80 MMS blankets and 48 divertor cassettes with a combined mass of ~3500t. The activated in-vessel components will require active cooling for a period of up to 18 months. This allows the decay heating to reduce so that the components have a stable temperature of ~50°C without additional cooling. This temperature is considered suitable for dextrous remote handling operations. The size of the component storage areas will be fairly constant for most maintenance scenarios (see § 5).

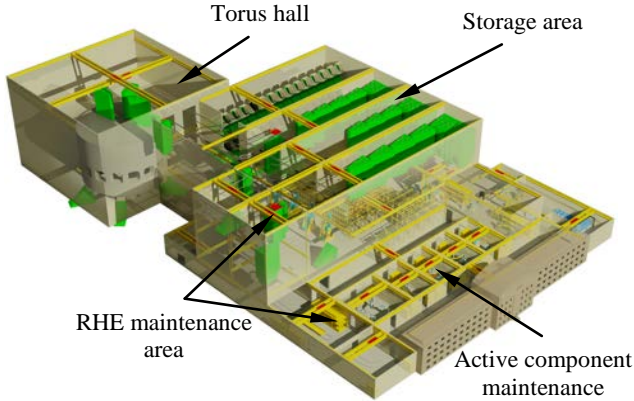

Fig. 4. Active Maintenance Facility

To facilitate replacement of the breeding blanket in 6 month, see § 5, the remote handling equipment storage areas will be extensive at ~140,000m$^3$ and will need to accommodate multiple casks of each design, totalling 35-40 casks. The casks for the blanket have an average mass of 350t and 150t for the divertor. This gives a combined total of ~ 11,000t.

The component maintenance areas are designed for specific components and are divided up into cells. These cells are capable of handling one blanket module or two divertor cassettes, see figure 5. Each cell can be maintained or reconfigured without affecting the others providing significant throughput benefits.

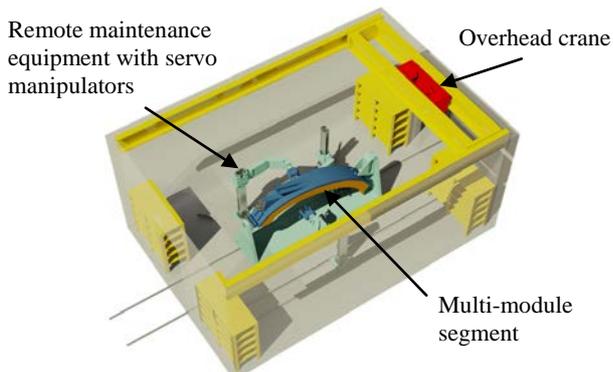

Fig. 5 MMS blanket maintenance cell

The maintenance of the remote handling equipment will be carried out in the AMF. The equipment will pass through a rigorous fully remote decontamination process in a dedicated cell before moving to an area where manual operations can be carried out.

The AMF will require many remote handling systems to be operated in parallel. During shutdowns the AMF operation will be focused on supplying and receiving components and equipment. Only items that will be returned to the machine during the shutdown will be processed in the AMF. Between shutdowns the AMF will process the in-vessel components and prepare the remote handling equipment for the next maintenance campaign. All of the work must be completed before the next shutdown. Splitting the operating modes will allow the operators to be moved between tasks allow a more consistently sized and skilled workforce.

The AMF must have sufficient capacity to avoid the shutdown critical path. It is highly challenging to meet the power plant availability, so it is essential that in-vessel operations are not delayed by AMF throughput.

## 5. Remote maintenance duration

The maintenance duration for the remote replacement of the shielding and tritium breeding blankets and the divertor cassettes has been estimated for a power plant [18]. The estimate is based on the EFDA DEMO pre-conceptual design studies for 2012 [9, 14], and uses data extrapolated from recorded times and operational experience from the remote maintenance activities on the JET Tokamak [19, 20].

The results suggest that for a highly developed and tested maintenance system with a large element of parallel working and with challenging but feasible operation times, the replacement of the blanket and divertor components could be achieved within the desired time frame of 6 months.

### 5.1 Basis of the estimate

The maintenance duration is estimated using a bottom-up approach, by summing the operation time for each task and then multiplying this by a number of factors:

A shift productivity ratio was used to take into account the number of hours of remote maintenance operation carried out each day. The estimate assumed a two shift operation pattern which provides 16 hours per day. This is a ratio of 0.67 and requires three shift teams.

An operator productivity factor was used to reflect the average rate of actual remote task completion against the theoretical maximum provided by the plan. The estimate assumed that automated processes such as cask docking or blanket extraction have a factor of 90% and a factor of 70% was used for the man-in-the-loop processes involving dexterous manipulation such as deploying tools.

Delays due to crane utilization were calculated but added less than 1% to the total duration due to the low utilization and relatively short duration of each crane operation.

Delays due to equipment failure were also calculated. Estimates were made for both equipment and process reliability and the time to recover from each type of failure was estimated. The delays due to remote handling equipment failure are approximately 20% but this is highly dependent on the reliability figures assumed. An intrinsic availability of 80% is high for such a complex system but is due to redundant or spare systems being provided wherever possible, the relatively high assumed reliability for the highly developed and tested system and because it is assumed that replacement casks are available on demand from the AMF to replace failed equipment thereby allowing rapid recovery and return to operations.

### 5.2 Assumptions

The remote handling equipment has been highly developed and tested and is operated by skilled and experienced staff leading to high productivity and reliability.

16 upper vertical ports and 16 lower divertor ports are available for remote maintenance and allow for simultaneous operations.

The AMF, see § 4, is capable of supplying and receiving casks without delaying remote operations with a suitable supply of spares to account for all likely failure scenarios.

The upper and lower port cask movers are capable of a cask move in half an hour.

Bore welding is used as the pipe joining technology with a triple weld head employed on large pipes. 2% of welds fail to meet the acceptance criteria after helium leak and volumetric testing.

Health physics checks are conducted using remote instrumentation allowing approval to be given from readouts at the operators control station.

### 5.3 Results from the estimate

The results produced by this estimate require four remote maintenance systems to operate in parallel to replace the blankets and cassettes within 6 months or to replace just the cassettes in 4 months. However, it is important to consider the accuracy of the input data when drawing conclusions from the estimate.

The final design of a power plant will require additional remote maintenance operations that are not included in the estimate and the duration and reliability figures used are informed estimates extrapolated from the most relevant existing systems but there are no systems as complex and highly developed as will be required for a fusion power plant.

Therefore the maintenance duration results should be taken as a first indication only.

## 6 Conclusions

The studies indicate the remote maintenance of a pre-conceptual design of a power plant to meet the required availability is achievable through a VMS approach, despite the challenging environmental conditions. It is imperative that the proposed maintenance strategy is demonstrated through mock-up studies at an early stage to inform decisions as to the integrated power plant/DEMO design.

The AMF and the remote handling equipment, forms a significant part of the overall plant design and cost and must not be underestimated to avoid impinging on the plant availability.

The studies indicate that the dexterous remote operations required for opening and closing the ports accounts for about three quarters of the total maintenance duration. This highlights the need to keep the in-vessel designs simple and the parts count low. Parallel operation in this area is essential to meet the availability targets.

The estimate also forms a base line, against which the impact of changes to component designs can be assessed in terms of the effect on maintenance duration and thereby the availability of the power plant.

The DEMO reactor will not have the same cost driver to minimize down time compared to a power plant. DEMO may be able to demonstrate the power plant technology using less remote handling systems operating in parallel and thereby reduced cost.

The remote maintenance design will be further advanced at an increased rate in 2014-2018 as part of the Design and R&D Programme, defined to implement the new European Fusion Roadmap and in particular laying the foundation of a DEMO to follow ITER, with the capability of generating several hundred MW of net electricity to the grid and operating with a closed fuel-cycle by 2050 [21].

### Acknowledgments

This work is funded by the RCUK Energy Programme [grant number EP/I501045] and by EFDA. To obtain further information on the data and models underlying this paper please contact *PublicationsManager@ccfe.ac.uk*. The views and opinions expressed herein do not necessarily reflect those of the European Commission.